\documentclass[a4paper,12pt]{article}
\usepackage[utf8]{inputenc}
\usepackage{cancel}
\usepackage{ulem}
\usepackage{amsfonts}
\usepackage{amssymb}
\usepackage{graphicx}
\usepackage{amsmath}
\usepackage{enumerate}
\usepackage{mathtools}
\usepackage{subfig}
\usepackage{color}
\usepackage{tikz}\usetikzlibrary{calc}
\usepackage{float}
\usepackage{here}
\usepackage{cite}
\usepackage{mathrsfs}
\usepackage{float,epsfig}
\usepackage{dcolumn}
\usepackage{tabularx}
\usepackage{multirow}
\usepackage{graphicx}
\usepackage{bm}
\usepackage{amsmath,amssymb,amsthm}
\usepackage[colorlinks=true,linkcolor=blue,citecolor=red]{hyperref}
\newcommand{\be}{\begin{equation}}
\newcommand{\ee}{\end{equation}}
\newcommand{\bea}{\setlength\arraycolsep{2pt} \begin{eqnarray}}
\newcommand{\eea}{\end{eqnarray}}

\def\0{{\sst{(0)}}}
\def\1{{\sst{(1)}}}
\def\2{{\sst{(2)}}}
\def\3{{\sst{(3)}}}
\def\4{{\sst{(4)}}}
\def\5{{\sst{(5)}}}
\def\6{{\sst{(6)}}}
\def\7{{\sst{(7)}}}
\def\8{{\sst{(8)}}}
\def\sst#1{{\scriptscriptstyle #1}}

\usepackage{pgfplots}

\makeatletter \@addtoreset{equation}{section}

\usepackage{tikz}
\usepackage{tkz-graph}

\usepackage{multirow}
\usepackage{tikz}

\begin{document}
%

\title{\normalsize
{\bf \Large	  On  Folding Calabi-Yau Diagrams in M-theory Black Brane Scenarios\footnote{This work is dedicated to the memory of   Fatima Addoud,  mother of the  first author.} 
 }}
\author{ \small Adil Belhaj,   Abderrahim Bouhouch\footnote{Authors  are listed in alphabetical order.}
	\hspace*{-8pt} \\
	{\small   ESMaR, Faculty of Science, Mohammed V University in Rabat,  
	Rabat, Morocco }\\
	}

 \maketitle

	\begin{abstract}
{\noindent}   
In this paper,  we  reconsider the study  of  five-dimensional supersymmetric black branes  in the context of  the  M-theory  compactification on  a special  Calabi-Yau manifold  called   tetra-quadric,  being   realized as  complete intersections  of  homogeneous polynomials  in  the projective space $ \mathbb{CP}^{1}\times\mathbb{CP}^{1}\times\mathbb{CP}^{1}\times\mathbb{CP}^{1}$.   Combining  colored graph theory with  outer-automorphism  group action techniques,  we approach   the   tetra-quadric Calabi-Yau  diagram leading to new features.  Using a  procedure referred to as folding, we show  that  M-theory  black branes on the   tetra-quadric  Calabi-Yau manifold  
 can be  reduced to  known compactifications  with lower  dimensional   K\"{a}hler moduli 
spaces.

\end{abstract}
 \newpage
\tableofcontents

%

\newpage

\newpage
\section{Introduction }

The construction of  five-dimensional (5D)  supersymmetric black branes has attracted a lot of attention and has been considered from different angles in the context of the compactification of M-theory on Calabi-Yau (CY) manifolds. The approach of interest here is the  derivation  of  the  black holes and the black  strings   using  the 5D $\mathcal{N}=2$
supergravity formalism \cite{1,2,3,4,5,6,7,70,71,72,73}.     BPS and  non-BPS 
   states  have been obtained   by  considering     M-branes  wrapping  on  non-holomorphic cycles   of the CY threefolds by the  help of  intersecting  number  calculations.    These calculations   depend on a 
real number $h^{1,1}$  that is  the K\"{a}hler moduli
space dimension of the  CY threefolds. Only lower dimensional cases have been approached using various methods including the analytical and numerical ones \cite{1,2,3,4,5,6,7}. 
  Two different CY  geometries  have been investigated.    Concretely,  a toric  geometry description   of M-theory scenarios   has  been largely  studied.   Certain  calculations   for   such CY threefolds regarded    as  hypersurfaces in toric varieties (THCY)    with   $h^{1,1}=3$ and $h^{1,1}=4$  have been provided \cite{3,6,7}.  In these investigations, several    5D
BPS and   non-BPS black  brane   configurations  involving stable and unstable
behaviors have been  derived   using numerical techniques\cite{3}. Alternatively,      
geometries   as   complete intersection CY threefolds  (CICY's)  in products of projective spaces  have been also   studied    via the    5D $\mathcal{N}=2$ supergravity 
formalism  \cite{8,9,10,11,12,13,14,140,141}.  Precisely,   a  M-theory CY threefold  in  the 
 $\mathbb{P}^{1}\times \mathbb{P}^{1}\times \mathbb{P}^{2}$ projective space product  has been  investigated  by  calculating  the corresponding  effective potential.  In this  model,    5D BPS and non-BPS black  brane   solutions have been  analyzed.   
 Stable and unstable  states    depending   on  the  charge
regions of the   K\"{a}hler moduli space have been determined using analytical and numerical computations \cite{6}. These discussions have been  elaborated   by evaluating  a  scalar quantity
called the recombination factor $R$.    It has been suggested that stable and unstable black  objects  are associated with    $R<1$ and $R>1$, respectively \cite {1}.

 In connection with black holes \cite{AAA}, graph theory has been exploited to make contact with quibit information models. In such an interplay,   vertices and edges have been used to encode many physical  data.  Moreover, graph theory  has been applied to  many problems including  the the minimum vertex cover (VC) and  
correcting codes  \cite{2500,2501}.

 The objective  of the present  paper  is to contribute to the program of    the   construction  of 5D  supersymmetric black branes using the  M-theory  compactification on  a special CY called   tetra-quadric CY.   Concretely,  we  reconsider the study  of   5D  supersymmetric black branes using such a CY,  being   realized as  complete intersections  of   homogeneous  polynomials  in  the projective space $ \mathbb{CP}^{1}\times\mathbb{CP}^{1}\times\mathbb{CP}^{1}\times\mathbb{CP}^{1}$.    Precisely,  we approach   the   tetra-quadric CY  diagram  providing new features by combining  colored graph theory and outer-automorphism  group action techniques. Using a  procedure referred to as folding and the scalar potential computations, we reveal   that  M-theory  black branes on the   tetra-quadric  CY  manifold  
 can be  reduced to  known  compactifications  with lower  dimensional   K\"{a}hler moduli 
spaces.

The organization of this paper is as follows. In section 2, we elaborate 
a  concise discussion  on  CICY manifolds by introducing a new procedure  in the CY diagrams using  colored  graph theory techniques.  In
section 3, we compute the effective scalar potential  of  black branes   from M-theory on the  tetra-quadric CY.    In section 4,
 we show  that  M-theory  black branes on the   tetra-quadric CY
 can be  reduced to  known  compactifications  with lower  dimensional   K\"{a}hler moduli  using the folding techniques. Section 5 contains
 concluding remarks.
\section{On    complete intersection Calabi-Yau threefolds }
In this section, we reconsider the study of certain features of CY manifolds known by CICY's. These manifolds have been extensively studied in superstring model compactifications and related topics  including M and F-theories  \cite{15,16,17,18,19}.  They are given by complete intersections  of    homogeneous  polynomials in a product of $m$  ordinary projective spaces. This product is described by an  ambient space  taking a general form given by
\begin{equation}
\mathcal{A}=\mathbb{CP}^{n_{1}}\times ...\times \mathbb{CP}^{n_{m}}
\end{equation}
where the involved integers $n_ i$, with $i=1,\ldots, m$,  can be fixed  by the CY  model in question. It is useful to recall that  a   $n_i$-dimensional ordinary projective space $\mathbb{CP}^{n_{i}}$ is defined by  the  following  scale identification 
\begin{equation}
z_\ell\sim \lambda z_{\ell},  \qquad \ell=1,\ldots, n_i+1
\end{equation}%
where $(z_{1},\ldots ,z_{n_{i}+1})$ are the   homogeneous 
coordinates  of  $\mathbb{CP}^{n_{i}}$ and   $\lambda $  denotes  a non-zero complex  number. In the present work,  we are interested in the  three-dimensional  $\mbox{CY}_3$  geometries  which can be
considered as CICY's in the  ambient
space $\mathcal{A}$. In this way,  certain constraints on   $n_{i }$  should be imposed.
\subsection{CY matrix configurations}
A close examination  shows that each  CICY can be represented by a matrix  configuration   carrying the most important data that are relevant in certain physical applications  such as high energy physics and related topics including black branes  in M-theory  compactification  scenarios \cite{1,2,3,4,5,6}. This matrix provides primordial information including the  geometric Hodge numbers 
$(h^{1,1}, h^{2,1})$  and  the Euler characteristic  $\chi$  being  a topological invariant.  Concretely,   it encodes features of the ambient space and the   homogeneous  degrees of the intersecting polynomials  needed to construct  CICY's.  The latters  are associated with the vanishing conditions of the involved   homogeneous   polynomials.   In  CICY theory,    each $\mbox{CY}_3$   can be represented by   a $m\times k$  integer matrix according to the following configuration 
\begin{equation}
\label{matrix}
	\mbox{CY}_3(\mathcal{A})= \left[\begin{array}{cccc}
	\mathbb{CP}^{n_{1}} \\ 
	\vdots \\
	\mathbb{CP}^{n_{m}}
	\end{array} \right|\left| \begin{array}{ccc}
	d_{1}^{1}& \ldots& d_{k}^{1}\\ 
	\vdots& \ddots & \vdots \\ 
	d_{1}^{m}& \ldots & d_{k}^{m}
	\end{array} \right]_{\chi}^{h^{1,1}, \; h^{2,1}}
	\end{equation}
such that 
\begin{equation}
\sum\limits_{i=1}^{h^{1,1}}n_{i}-k=3,\qquad n_{i}+1=\sum\limits_{r
=1}^{k}d^{i }_{r}
\end{equation}%
as required by the CY condition.  Several  examples of such $\mbox{CY}_3$ geometries have  been largely  investigated  via various classifications using different techniques and methods including  toric geometry.   A key observation shows that the most needed  quantities being important in string theory compactifications and  the black branes in M-theory are the triple intersection numbers  $C_{ijk}$, with  $1 \leq i, j, k \leq  h^{1,1},$  where one has identified $m$ with $h^{1,1}$.   These intersection  numbers  can be determined  using  the    K\"{a}hler  forms   ${\cal J}_1,\ldots,$  and    ${\cal J}_{h^{1,1}}$
via the relation 
\begin{equation}
 C_{ijk}= \int _{\cal A} \mu  \wedge  {\cal J}_i   \wedge  {\cal J} _j   \wedge   {\cal J} _k
\end{equation}
where  $\mu $ is a real  ($2\sum\limits_{i=1}^{h^{1,1}} n_i-6$)-form  expressed as follows 
\begin{equation}
 \mu=  \sum_{i=1}^m  (d_{1}^{i}  {\cal J}_i)    \wedge \ldots    \wedge   (d_{k}^{i}  {\cal J}_i).
\end{equation}
This algorithm leads  to the $\mbox{CY}_3$  volume  given by 
\begin{equation}
{\cal V}= \frac{ C_{ijk}   t ^i  t^j   t^k }{3!}
\end{equation}
where $ t^i$ are    the  scalar moduli corresponding   the  K\"{a}hler   forms denoted by  $ {\cal J}_i$. 
\subsection{$\mbox{CY}$  diagrams}
\begin{figure}[!h]

\begin{center}
\includegraphics[scale=0.25]{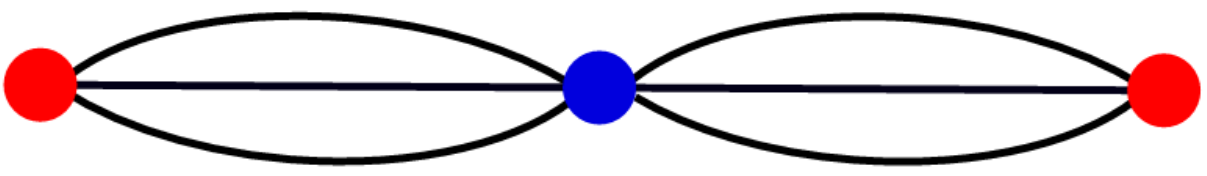}

\caption{ Diagram of bi-cubic   in $ \mathbb{CP}^{2}\times \mathbb{CP}^{2}$.}
\label{F1}
\end{center}
\end{figure}

In the constructions of  CICY's, we can observe certain similarities with the application of graph theory encoding the most relevant data either as Dynkin diagrams used in the classification of Lie algebras, or as Feynman diagrams exploited in quantum field theory calculations \cite{20,21,22}. According to \cite{23},  a  CICY manifold defined by the matrix configuration given by Eq.(\ref{matrix}) can be associated with   a diagram  denoted by  ${\cal D}$ according to the following scheme 
\begin{equation}
\mbox{CICY}  \to {\cal D}(\mbox{CY}_3)={\cal D}.
\end{equation}
This means that the matrix configuration    is completely encoded in the diagram $ {\cal D}$ shearing similarities  with the Cartan matrix and the   Dynkin diagrams of  the finite Lie algebras. As usually,  $ {\cal D}$  is a pair of  $(  V({\cal D}), E({\cal D}))$ where    $V({\cal D})$ and  $E({\cal D})$  denote the vertex and the  leg (edge) sets, respectively \cite{24,25,2500,2501}.  In fact, one distinguishes  two types of vertices associated with  rows and columns of the configuration matrix.  To make things more clear,  we  will consider diagrams involving  colors to reveal such  vertex distinguishable aspects describing the   homogenous polynomial    degrees    and the algebraic equation  constraints.  In such  $\mbox{CY}_3$  diagrams,   we consider two different colors producing diagrams with a chromatic number   equals  to 2. It is recalled that  this number,  in  graph theory, indicates the smallest number of colors required to color diagrams so that no two adjacent vertices have the same color.  To draw the $\mbox{CY}_3$   diagrams, one  can follow the steps below.   The red  color  symbolizes the vertices indicating  the ordinary projective space  factors and the  blue  one concerns  the vertices  representing  the algebraic equation constraints. Indeed, each  $ \mathbb{CP}^{n_{i}}$ factor  is represented by  a red vertex of degree $\sum\limits_{r
=1}^{k}d^{i }_{r}$  being identified  with the number  of outgoing legs
\begin{equation}
 \mathbb{CP}^{n_{i}} \to  v_i, \qquad  i=1,\ldots, h^{1,1}
\end{equation}
Via such legs, the red vertices  are linked to the blue ones   representing  the algebraic equation constraints $c_\alpha$.  This means that  each constraint  is represented by a blue vertex with  a degree  equals to   $\sum\limits_{i
=1}^{h^{1,1}}d^{i }_{r}$
\begin{equation}
c_r \to  v_r \qquad  r=1,\ldots, k.
\end{equation}
To show explicitly  such  a graphical method, we consider  the  bi-cubic in  the  projective space $ \mathbb{CP}^{2}\times \mathbb{CP}^{2}$, for instance,  where    Eq.(\ref{matrix}) reduces to 
\begin{equation}
\label{matrix1}
	\mbox{CY}_3(\mathcal{A})= \left[\begin{array}{cccc}
	\mathbb{CP}^{2} \\ 
	\mathbb{CP}^{2}
	\end{array} \right|\left| \begin{array}{ccc}
	3\\ 3
	\end{array} \right]_{\chi}^{h^{1,1}, \; h^{2,1}}
	\end{equation}
with  $n_1=n_2=2$ and  $d_{1}^{1}=d_{1}^{2}=3$. To construct its  diagram, we need two red vertices  of degree 3 and one blue of degree six.  This bi-cubic  diagram can be illustrated in Fig(\ref{F1}).

Having given  the most  relevant data of such  CY  building models via graph theory techniques, we move to 
provide new features  corresponding to discrete group actions.
	
\subsection{ Folding  CY diagrams }	
\begin{figure}[!h]
\begin{center}
\includegraphics[scale=0.25]{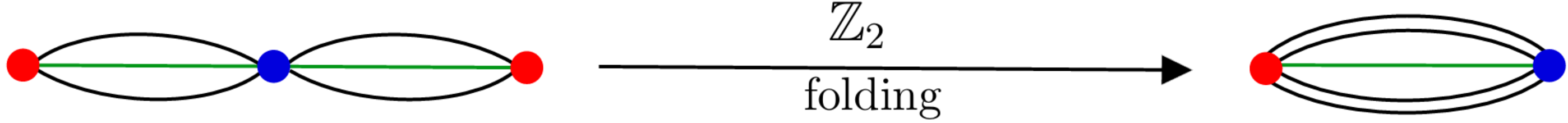}
\caption{  $\mathbb{Z}_{2}$ folding of diagram of bi-cubic   in $ \mathbb{CP}^{2}\times \mathbb{CP}^{2}$.}
\label{F2}
\end{center}

\end{figure}

Inspired by  Dynkin diagram techniques\cite{26,27}, we would like to generate a new feature from CY diagrams  using the so-called folding procedure.   This can be done with the help of an  outer-automorphism  group $\Gamma$ leaving the CY digram  invariant 
	\begin{equation}
\Gamma: {\cal D} \to  {\cal D}.
\end{equation}
The folding procedure can identify  vertices with  the same color and the same degree  which are permuted by the discrete group  $\Gamma$ being a subgroup of  the $\mathbb{S}_m \times \mathbb{S}_k$ permutation structure. This scenario provides a new diagram with certain reductions in the resulting  $\mbox{CY}_3$  diagrams by putting such vertices  in the same orbit.   For red vertices,  this group action results in  the  K\"{a}hler   moduli space by decreasing its dimension. This can produce  certain topological change in the folded   geometries. This  transition could find a  relevant place  in the elaboration of M-theory black branes on   CICY manifolds. 
To see how such a new procedure works, we consider a  toy model given by the  bi-cubic in  the $ \mathbb{CP}^{2}\times \mathbb{CP}^{2}$  projective space product. The corresponding diagram is invariant under the  $ \mathbb{Z}_{2}$ discrete symmetry. According to   Fig(\ref{F1}),  this group    permutes the red nodes and  leaves the blue one invariant.   These two red vertices are in the same orbit of such a   $ \mathbb{Z}_{2}$ symmetry. In fact, they transform as a doublet in the folding scenario language.   In this way, this geometric  procedure identifies  these  two  red vertices producing just one. A priori,  there are  many scenarios  which  may depend on the action of  the discrete  symmetry $\mathbb{Z}_{2}$ on the legs of the  $ \mbox{CY}_3$  diagrams.  To keep the right dimension of   CICY models, we identify just two green  legs to provide   only one in the folding procedure. After  such a  folding   action, we get the quintic CY diagram as illustrated in   Fig(\ref{F2}).

In what follows, we show that such folding diagrams of   CY  threefolds  can  be explored in  the study of  5D black branes from   the M-theory compactification on the  tetra-quadric  CY manifold. 

\section{M-theory  black branes  from the  tetra-quadric CY }
\begin{figure}[!h]
\begin{center}
\includegraphics[scale=0.25]{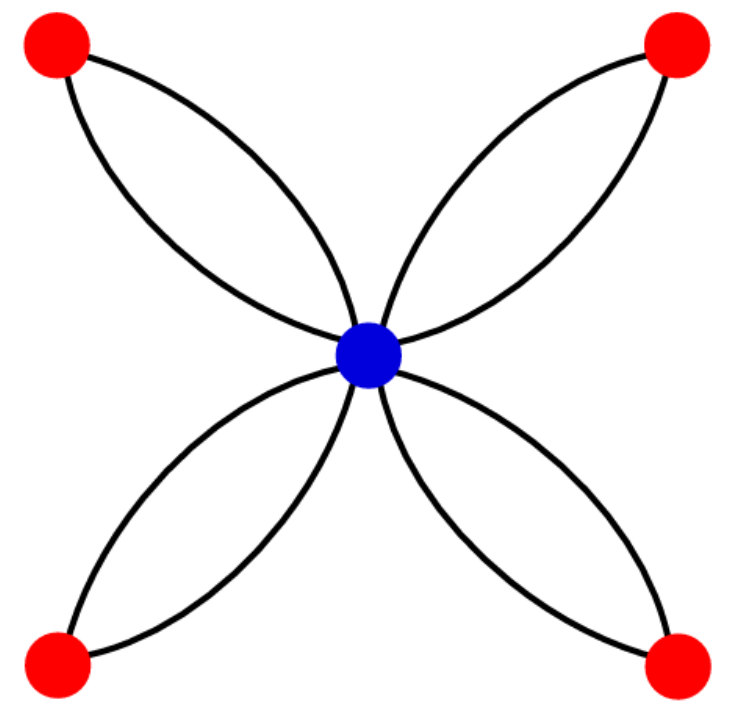}
\caption{  Tetra-quadric CY diagram. }
\label{F3}
\end{center}
\end{figure}

In this section,  we reconsider the investigation   of  5D  black holes and black  strings     from  the so-called favorable   $\mbox{CY}_3$  which is a tetra-quadric  CY \cite{1}.   The manifold  is embedded
in the following  ambient space
\begin{equation}
\mathcal{A}=\mathbb{CP}^{1}\times\mathbb{CP}^{1}\times\mathbb{CP}^{1}\times\mathbb{CP}^{1}
\end{equation}
being  a product of four  $\mathbb{CP}^{1}$ ordinary projective  spaces. It can be  defined  as the zero locus of  quadratic homogeneous   polynomials of degree $(2,2,2,2)$  
in the homogeneous coordinates of  $\mathcal{A}$  having  the following geometric and topological data
\begin{equation}
(h^{1,1}, h^{2,1})=(4,68), \qquad \chi=-128.
\end{equation}
All   these  data can be encoded in the following configuration matrix
\begin{equation}
\mathcal{A}:=\left[\begin{array}{ccc}
\mathbb{CP}^{1} \\ 
\mathbb{CP}^{1}\\ 
\mathbb{CP}^{1}\\ 
\mathbb{CP}^{1}
\end{array} \right|\left| \begin{array}{c}
2\\ 
2\\
2\\
2
\end{array} \right]_{-128}^{4,68}.
\end{equation}
According to  graph theory language \cite{24,25,2500,2501}, the diagram of such a manifold  involves  four red vertices of degree 2 linked to a blue one of degree 8. It is illustrated  in Fig(\ref{F3}).

The primordial  geometric data of  such  a CY are   the  intersection numbers being useful    to determine the  relevant   quantities of  black  branes  in M-theory  including the effective  scalar potential needed to approach  certain physical    behaviors such as stability. Indeed, the triple intersection numbers are  found to be 
\begin{eqnarray}
C_{123}&=&C_{124}= C_{134}=C_{234}=2\\
C_{ijk}&=&0 \,\,\,\text{if },\,\,\,\,i=j=k,\,\,i\neq j=k,\,\,\  i\neq j=k,\,\,\  \text{or }\,\, i=k\neq j.
\end{eqnarray}
These intersection numbers provide the volume of  the proposed CY being expressed  as follows
\begin{equation}
\mathcal{V}=2 t_1 t_2 t_3+2 t_1 t_3 t_4+2 t_2 t_3 t_4+2 t_1 t_2 t_4
\end{equation}
linked to 
the   K\"{a}hler  moduli space metric $G_{ij}$   via the relation
\begin{equation}
G_{ij}=-\frac{1}{2}\partial _{i}\partial _{j}\log (\mathcal{V}).
\end{equation}%
The  M-theory compactification on such a CY   can be approached   by  the help of  5D $\mathcal{N}=2$  supergravity formalism  \cite{8,9,10,11,12,13,14}. 
In this regard,  the M-theory  black branes  can be constructed by exploiting the   M2 and
M5-branes wrapping on non-trivial cycles in   the tetra-quadric CY corresponding to the  $ U(1)^{\times 4}$ gauge symmetry.   Indeed, these objects can  be  dealt with  via  the following 5D   Maxwell-Einstein action 
\begin{equation}
S=\frac{1}{2\kappa _{5}^{2}}\int d^{5}x\left( R\star \mathbb{I}%
-G_{ij}dt^{i}\wedge \star dt^{j}-G_{ij}F^{i}\wedge \star F^{j}-\frac{1}{6}%
C_{ijk}F^{i}\wedge F^{j}\wedge A^{k}\right) 
\end{equation}%
where   $t_{i}$ are   the  scalar   K\"{a}hler  moduli  and  $F^{i}=dA^{i}$ denote  the Maxwell  gauge fields.   Roughly, the geometric quantities    $C_{ijk}$  and $G_{ij}$      being the intersecting numbers and 
the   K\"{a}hler  moduli space metric, respectively, will be   needed to  calculate    the
effective scalar potential of   the   black branes in the M-theory compactifications on the  tetra-quadric CY manifold.   
\subsection{Black holes }
5D black holes in  the M-theory on the  tetra-quadric CY involve four electric  charges  $(q_1,q_2,q_3,q_4)$  under the   $ U(1)^{\times 4}$ gauge symmetry.     In this  building solution, the electric  charges are associated with the  M2-branes wrapping  on  2-cycles in such a CY threefold. The corresponding  central charge   can  be expressed as
\begin{equation}
Z_{e}=q_1 t_1+ q_2 t_2+ q_3 t_3+q_4 t_4
\end{equation}
where one has used four  scalar  moduli $t_i$ satisfying  the  K\"{a}hler  cone conditions  $   t_i \geq 0 $,  $1\leq i\leq 4.$ 
 To approach certain  physical behaviors, one  should compute  the effective scalar  potential  of  such 5D black holes in M-theory scenarios. This can be done  using the relation \begin{equation}
\label{Vbh}
V_{eff}^{BH}(q_i,t_i)= G^{ij} q_iq_j, \qquad  i,j=1,\ldots, 4.
\end{equation}
Computations reveal that  this   scalar potential   is  found to  be 
\begin{equation}
V_{eff}^{BH}(q_i,t_i)=\frac{G(q_i,t_i)}{T(t_i)}
\end{equation}
where  $T$  is  a  geometric scalar  function  depending  only  on the   K\"{a}hler   moduli  given by 
\begin{equation}
T(t_i)=  \left(t_3 t_4+t_2 \left(t_3+t_4\right)\right) t_1^2+\left(\left(t_3+t_4\right) t_2^2+\left(t_3^2+t_4^2\right) t_2+t_3 t_4 \left(t_3+t_4\right)\right) t_1+t_2 t_3 t_4 \left(t_2+t_3+t_4\right).
\end{equation}
The  scalar   quantity  $G(q_i,t_i)$  can be expressed as follows 
\begin{equation}
G(q_i,t_i)=
g^{ij}(t_i)q_{i}q_{j}
\end{equation}
where one has used  the following matrix elements 
\begin{align*}
g^{11}&= 2 (t_3 t_4+t_2(t_3+t_4)) t_1^4+2 (t_2+t_3) (t_2+t_4) (t_3+t_4) t_1^3+((t_3^2+4 t_4 t_3+t_4^2) t_2^2+4 t_3 t_4 (t_3+t_4) t_2\\
&+t_3^2 t_4^2) t_1^2+2 t_2 t_3 t_4(t_3 t_4+t_2 (t_3+t_4)) t_1+t_2^2 t_3^2 t_4^2
\,\\
g^{12}&=(t_2^2 (t_3-t_4)^2-t_3^2 t_4^2) t_1^2-2 t_2 t_3^2 t_4^2 t_1-t_2^2 t_3^2 t_4^2
\,\\
g^{13}&=(t_1 (t_2 (t_3-t_4)-t_3 t_4)-t_2 t_3 t_4) (t_2 t_3 t_4+t_1(t_2(t_3+t_4)-t_3 t_4))
\,\\
g^{14}&=-(t_2 t_3 t_4+t_1 (t_2 (t_3-t_4)+t_3 t_4)) (t_2 t_3 t_4+t_1 (t_2 (t_3+t_4)-t_3 t_4))
\,\\
g^{22}&=(2 (t_3+t_4) t_2^3+(t_3^2+4 t_4 t_3+t_4^2) t_2^2+2 t_3 t_4 (t_3+t_4) t_2+t_3^2 t_4^2)t_1^2+2 t_2((t_3+t_4)t_2^3\\&+(t_3+t_4)^2 t_2^2+2 t_3 t_4 (t_3+t_4) t_2+t_3^2 t_4^2) t_1+t_2^2 t_3 t_4 (2 t_2^2+2 (t_3+t_4) t_2+t_3 t_4) 
\,\\
g^{23}&=(t_1 (t_2 (t_3-t_4)-t_3 t_4)-t_2 t_3 t_4) (t_1 (t_3 t_4+t_2 (t_3+t_4))-t_2 t_3 t_4)
\,\\
g^{24}&=-(t_1 (t_3 t_4+t_2 (t_3+t_4))-t_2 t_3 t_4) (t_2 t_3 t_4+t_1(t_2 (t_3-t_4)+t_3 t_4))
\,\\
g^{33}&= 4t_1 t_2 t_4^2 t_3^2+2 t_1 t_2 t_3^4+2 t_1 t_4 t_3^4+2 t_2 t_4 t_3^4+2 t_1 t_2^2 t_3^3+2 t_1 t_4^2 t_3^3+2 t_2 t_4^2 t_3^3+2 t_1^2 t_2 t_3^3+2 t_1^2 t_4 t_3^3\\
& +2 t_2^2 t_4 t_3^3+4 t_1 t_2 t_4 t_3^3+t_1^2 t_2^2 t_3^2+t_1^2 t_4^2 t_3^2 +t_2^2 t_4^2 t_3^2+4 t_1 t_2^2 t_4 t_3^2+4 t_1^2 t_2 t_4 t_3^2+2 t_1 t_2^2 t_4^2 t_3+2 t_1^2 t_2 t_4^2 t_3\\
&+2 t_1^2 t_2^2 t_4 t_3+t_1^2 t_2^2 t_4^2.
\end{align*}


\subsection{Black strings}
The compactification  of M-theory on   the  tetra-quadric CY   manifold can produce   also 5D black strings  with four magnetic charges  $(p_1,p_2,p_3,p_4)$  under the   $ U(1)^{\times 4}$ gauge symmetry.    These black   brane objects  can be built using   M5-branes wrapping on   4-cycles in  such a CY manifold. As  in the black hole case,  we  should   compute  the black string effective  potential needed to approach the  associated physical behaviors.  According to \cite{1},  the black string effective potential   $V^{BS}_{eff}$  can be determined via the relation  
\begin{equation}
\label{vm}
V^{BS}_{eff}= 4 G_{ij} p^ip^j.
\end{equation}
Computations lead to 
\begin{eqnarray*}
\label{vm}
V^{BS}_{eff}&=& 8 p_1^2 (t_2^2 t_3^2 + 2 t_2^2 t_3 t_4 + 2 t_2 t_3^2 t_4 + t_2^2 t_4^2 + 
2 t_2 t_3 t_4^2 + t_3^2 t_4^2)\\&&
+ 8p_2^2 ( t_1^2 t_3^2 + 2 t_1^2 t_3 t_4 + 2 t_1 t_3^2 t_4 + 2 t_1 t_3 t_4^2 + t_1^2 t_4^2 +
 t_3^2 t_4^2)
 \\&&+8p_3^2  (t_2 t_4 + t_1 (t_2 + t_4))^2
+8p_4^2  (t_2 t_3 + t_1 (t_2 + t_3))^2
 \\&&
+16 p_1p_2 t_3^2 t_4^2+16 p_1p_3 t_2^2 t_4^2+16 p_1p_4  t_2^2 t_3^2+16 p_2p_3  t_1^2 t_4^2+16 p_2p_4  t_1^2 t_2^2+16 p_3p_4 t_1^2 t_3^2.  
\end{eqnarray*}
 Having computed the 5D   black brane scalar  potentials,  we move to approach   the folding  tetra-quadric CY   diagram  in M-theory  compactification  scenarios. 
\section{Folding  tetra-quadric CY   diagram in   M-theory black brane  scenarios }

In this  section,  we  approach  the black brane 
physics   resulting from  the compactification of M-theory.  At first sight, such a M-theory  physics 
appears  quite complicated. However,  we will discuss how M-theory on the   tetra-quadric CY
can be  reduced to  M-theory  on CY 
threefolds with lower  dimensional   K\"{a}hler moduli spaces. This  link    relays  on a geometric
procedure called folding. In string theory combined with toric geometry, this
procedure has been  explored  to geometrically engineer non-simply-laced
 gauge theories using quiver techniques \cite{26,27}. In this procedure, one can  identify  the 
vertices of the CY threefold  diagrams  being permuted under a 
folding action $\Gamma$ considered as an outer-automorphism of the
associated   diagram $\cal D$. This imposes certain constraints on  the tetra-quadric CY data  depending on the precise action of $\Gamma$. The resulting   K\"{a}hler    geometries involve some dimensions less than the natural
one. This dimensional reduction follows straightforwardly from the  K\"{a}hler moduli space behaviors. Indeed, the folded resulting  diagrams can be obtained from  the  tetra-quadric CY diagrams by identifying  red vertices and green legs   which are permuted by the outer-automorphism  group  $\Gamma$. It follows from the  tetra-quadric CY diagram that  the non-trivial groups leaving such a diagram invariant are
\begin{eqnarray}
\Gamma= \mathbb{Z}_{2}, \quad  \mathbb{Z}_{2} \times \mathbb{Z}_{2}, \quad   \mathbb{Z}_{3}, \quad   \mathbb{Z}_{4}.
\end{eqnarray}
being listed in Tab.(\ref{T1}).
\newpage
\begin{table}[!h]
\begin{center}
\includegraphics[scale=0.95]{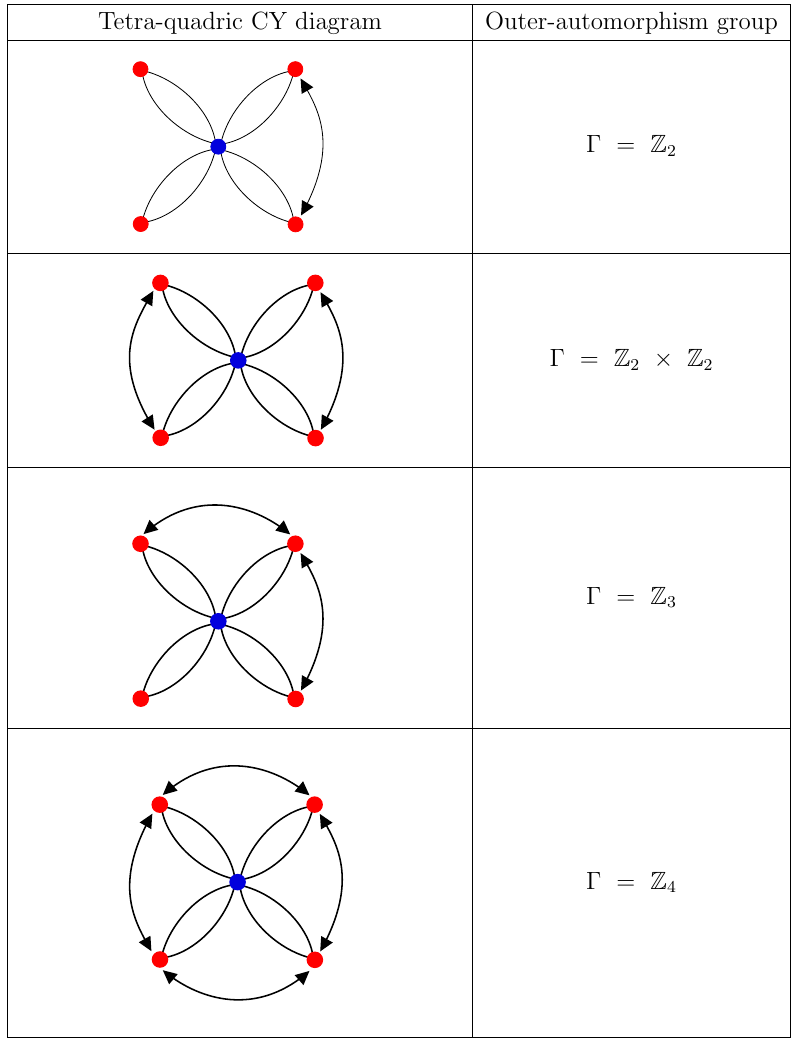}
\caption{Tetra-quadric CY diagram and its outer-automorphism groups}
\label{T1}
\end{center}
\end{table}

\newpage

\subsection{$ \mathbb{Z}_{2}$ folding  procedure  in M-theory scenario}
\begin{figure}[!h]
\begin{center}

\includegraphics[scale=0.25]{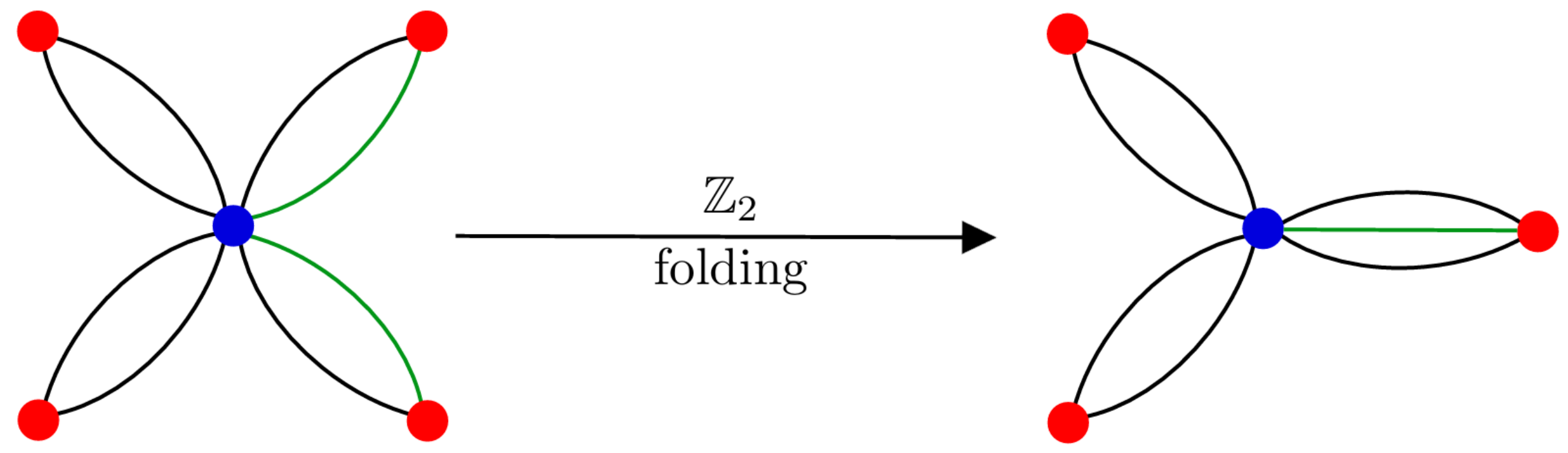}
\caption{ $ \mathbb{Z}_{2}$  folding of  tetra-quadric CY diagram. }
\label{F4}
\end{center}

\end{figure}

To understand  the folding procedure,  we   first consider  the  $ \mathbb{Z}_{2}$ group action identifying two red vertices associated with two different $ \mathbb{CP}^{1}$'s in the  tetra-quadric CY configuration. In this way, these two vertices are in the same orbit of  the $ \mathbb{Z}_{2}$   reflection  symmetry.  They transform as a doublet. The folding scenario identifies  these vertices and the associated green links   as  required by  the CY  dimension condition. This scenario is illustrated in  Fig.(\ref{F4}).
This folding scenario leads to a digram which can be  associated with  a CY threefold in the projective space   $\mathbb{CP}^{1}\times\mathbb{CP}^{1}\times\mathbb{CP}^{2}$.   The resulting  K\"{a}hler geometry involves one dimension less than  the natural one. This dimension reduction can be interpreted as a truncation  in  the black hole physics embedded in  M-theory on the  tetra-quadric CY. It has  been observed that the  $\mathbb{Z}_{2}$  action  can be consorted by certain constraints that should  be imposed in the  K\"{a}hler  and  the charge moduli spaces.
On such spaces, the  $\mathbb{Z}_{2}$  action  can be accompanied by the following transformations  on the $t_i$ and $q_i$ quantities 
 \begin{eqnarray}
 &&\left(t_1, t_2, t_3, t_4\right) \to  \frac{1}{3} \sqrt{\frac{2}{3} } \left(t_1, t_2,  \frac{3}{2}t_3,  \frac{3}{2}t_3\right)\\
&& \left(q_1, q_2, q_3, q_4\right)  \to    3\sqrt{\frac{3}{2}} \left(   q_1,  q_2,\frac{4}{3}q_3, \frac{4}{3}q_3\right).
  \end{eqnarray}
Putting such  transformed   K\"{a}hler   moduli and charges in the   black hole scalar   potential Eq.(\ref{Vbh}), we recover  the scalar potential of 5D black holes obtained from M-theory on a   CICY  in  the projective space  $\mathbb{CP}^{1}\times\mathbb{CP}^{1}\times\mathbb{CP}^{2}$  reported in \cite{6}. In this way, 
the  black hole  effective potential  reduces to 
\begin{equation}
V^{BH}_{eff}= \frac{G(t_1,t_2,t_3,q_1,q_2,q_3)}{{ T(t_1,t_2,t_3)}},
\end{equation}
where  we have used 
\begin{align*}G(t_1,t_2,t_3,q_1,q_2,q_3)&=6 q_2 q_3 t_3^2 \left( t_2 t_3 - t_1 (3 t_2 + t_3)+ 3 q_3^2 t_3^2 (3 t_1 (3 t_2 + t_3)+ t_3 (3 t_2 + 2 t_3)\right)  \\
&+ q_1^2 \left( 2 t_2 t_3^3 + 9 t_1^3(3 t_2 + t_3)+ 2 t_1 t_3^2 (6 t_2 + t_3) + 3 t_1^2 t_3 (9 t_2 + 4 t_3)\right)  \\
&- 2 q_1 t_3^2 \left( 2 q_2 (t_1 + t_2) t_3 + 3 q_3 (3 t_1 t_2 - t_1 t_3 + t_2 t_3)\right) \\
&+ q_2^2 \left( t_2 t_3 (9 t_2^2 + 12 t_2 t_3 + 2 t_3^2) + t_1 (27 t_2^3 + 27 t_2^2 t_3 + 12 t_2 t_3^2 + 2 t_3^3)\right)\\
T(t_1,t_2,t_3) &=\frac{9 t_1 (3 t_2 + t_3)+ 3 t_3 (3 t_2 + 4 t_3)}{{2}}.
\end{align*} 
$$$$
 Similar transformations can be  provided  for  the   5D black string potential.   For such solutions,  the  $\mathbb{Z}_{2}$  actions  can be accompanied by the following transformations  on  $t_i$ and $p_i$    quantities 
 \begin{eqnarray}
 &&\left(t_1, t_2, t_3, t_4\right) \to \frac{1}{3}  \left(2 t_1,2  t_2, 3t_3,  3t_3\right)\\
&& \left(p_1, p_2, p_3, p_4\right)  \to    6\left(2p_1, 2p_2, 3p_3,  3p_3\right).
  \end{eqnarray}
Putting these  transformed   $t_i$ and $p_i$ quantities,   we recover  the scalar  potential of 5D black  strings  obtained from M-theory on  a     CICY  in  the projective space  $\mathbb{CP}^{1}\times\mathbb{CP}^{1}\times\mathbb{CP}^{2}$ investigated  in  \cite{6}.
 Using  similar techniques, the stringy 
 effective potential   can be reduced to 
\begin{eqnarray}
V^{BS}_{eff}&=& \frac{1}{18} (p_3^2 (9 t_1^2 t_2^2+6 t_1 (t_1+t_2) t_3 t_2+2 (t_1+t_2)^2 t_3^2)+6 p_3 t_3^2(p_2 t_1^2+p_1 t_2^2)\\
&&+t_3^2 (2 p_1 p_2 t_3^2+p_2^2 (3 t_1+t_3)^2+p_1^2 (3 t_2+t_3)^2)).\nonumber
\end{eqnarray}

\subsection{   $\mathbb{Z}_{2}\times \mathbb{Z}_{2}$   folding  procedure in M-theory scenario}
\begin{figure}[!h]
\begin{center}

\includegraphics[scale=0.25]{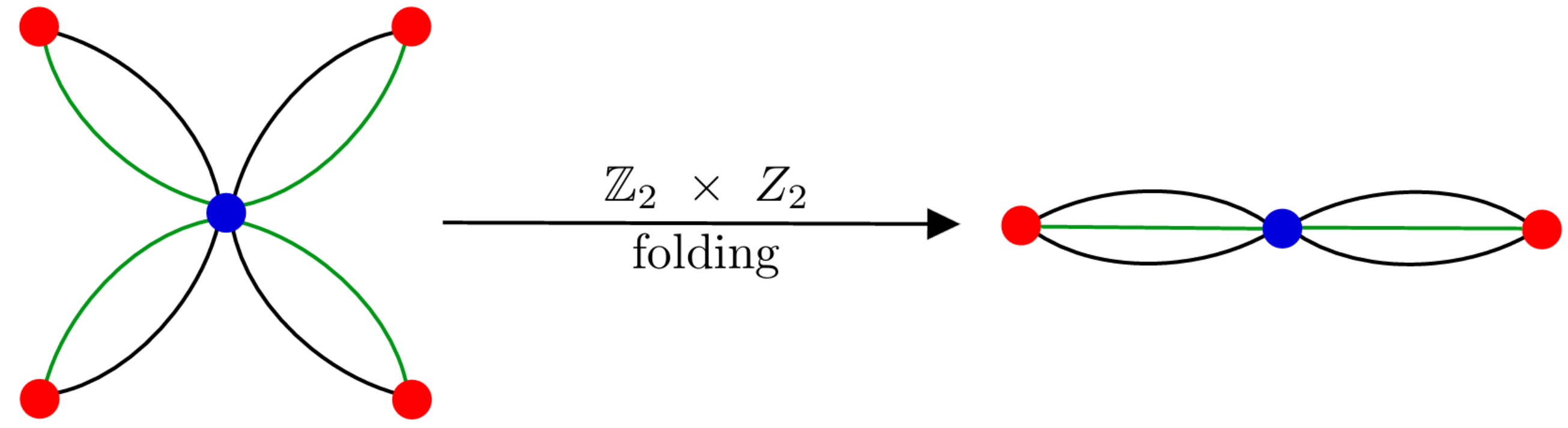}
\caption{ $ \mathbb{Z}_{2} \times  \mathbb{Z}_{2} $  folding of  the  tetra-quadric CY diagram. }
\label{F5}
\end{center}
\end{figure}

Following the same method of the folding scenario, we can  elaborate   the  $\mathbb{Z}_{2}\times \mathbb{Z}_{2}$   action    considered as an extended  $\mathbb{Z}_{2}$ symmetry. Indeed,  the first    $\mathbb{Z}_{2}$
 group  identifies   two red vertices associated with two different $ \mathbb{CP}^{1}$'s.   The second one  identifies  two red vertices of the remaining  two $ \mathbb{CP}^{1}$'s.  In this way,   these red vertices  transform as  two different  doublets associated with  two different orbits of  the  $\mathbb{Z}_{2}\times \mathbb{Z}_{2}$  symmetry.  As the previous scenario, this $\mathbb{Z}_{2}\times \mathbb{Z}_{2}$   folding scenario identifies  these  doublet vertices and the  corresponding  green links   as  required by  the CY  dimension condition. The  procedure is  shown  in  Fig.(\ref{F5}).

This folding scenario leads to a diagram which can  represent  the   bi-cubic in   the projective space    $\mathbb{CP}^{2}\times\mathbb{CP}^{2}$.   The resulting  K\"{a}hler geometry involves two  dimension less than  the natural one. This dimension reduction can be interpreted as a truncation  in the  black hole physics embedded in   the M-theory on the tetra-quadric CY. It has  been observed that the  $ \mathbb{Z}_{2} \times  \mathbb{Z}_{2} $ actions are joined  by  certain constraints that should  be imposed on  the  K\"{a}hler  and  the charge moduli spaces of the  M-theory on  the tetra-quadric CY. 
On such  moduli spaces, the   $ \mathbb{Z}_{2} \times  \mathbb{Z}_{2} $  actions   can be accompanied by the following transformations on  the $t_i$ and $q_i$  physical quantities 
 \begin{eqnarray}
 &&\left(t_1, t_2, t_3, t_4\right) \to  \sqrt{2} \left( t_1,t_1,  t_2, t_2\right)\\
&& \left(q_1, q_2, q_3, q_4\right)  \to    \sqrt{2} \left(  q_1,  q_1, q_2, q_2\right).
  \end{eqnarray}
Putting such  transformed   K\"{a}hler   moduli and charges in the   black hole scalar   potential Eq.(\ref{Vbh}), we recover  the scalar potential of 5D black holes obtained from M-theory on the bi-cubic in  the projective space  $\mathbb{CP}^{2}\times\mathbb{CP}^{2}$  reported in \cite{1}. In this way, 
the  black hole effective potential    of the M-theory on the  tetra-quadric CY reduces to 
\begin{equation}
V^{BH}_{eff}= \frac{G(t_1,t_2,q_1,q_2)}{{ T(t_1,t_2)}},
\end{equation}
where  we have used 
\begin{align*}G(t_1,t_2,q_1,q_2)&= q_1^2 t_1^2 \left(t_1^2+2 t_1 t_2+2 t_2^2\right)- 2q_1 q_2 t_1^2 t_2^2  +q_2^2t_2^2 \left(2t_1^2+2 t_1 t_2+ t_2^2\right) \\
T(t_1,t_2) &= t_1^2+t_1 t_2+t_2 ^2.
\end{align*} 
 Similar discussions can be elaborated  for black string potentials. Concerning the  solutions,  the  $ \mathbb{Z}_{2} \times  \mathbb{Z}_{2} $ actions  can be accompanied by the following transformations  on  $t_i$ and $p_i$   physical   quantities 
 \begin{eqnarray}
 &&\left(t_1, t_2, t_3, t_4\right) \to   \frac{2}{3}\left( t_1,  t_1, t_2,  t_2\right)\\
&& \left(p_1, p_2, p_3, p_4\right)  \to   6 \left(p_1, p_1, p_2,  p_2\right). 
  \end{eqnarray}
Considering such transformed   $t_i$ and $p_i$ physical variables,   we recover  the scalar  potential of 5D black  strings  obtained from M-theory on  the   bi-cubic in    the  ambient projective space   $\mathbb{CP}^{2}\times\mathbb{CP}^{2}$ investigated  in  \cite{1}.
 Using  similar techniques, this  stringy 
 effective potential  corresponding to    the   tetra-quadric CY  manifold  reduces  to 
\begin{eqnarray}
V^{BS}_{eff}= \frac{9}{2} \left(2 p_1 p_2 t_2^2 t_1^2+p_2^2 \left(t_1^2+2 t_2 t_1+2 t_2^2\right) t_1^2+p_1^2 t_2^2 \left(2 t_1^2+2 t_2 t_1+t_2^2\right)\right).
\end{eqnarray}

\subsection{  $\mathbb{Z}_{3}$  folding  procedure in M-theory scenario}

\begin{figure}[!h]
\begin{center}
\includegraphics[scale=0.25]{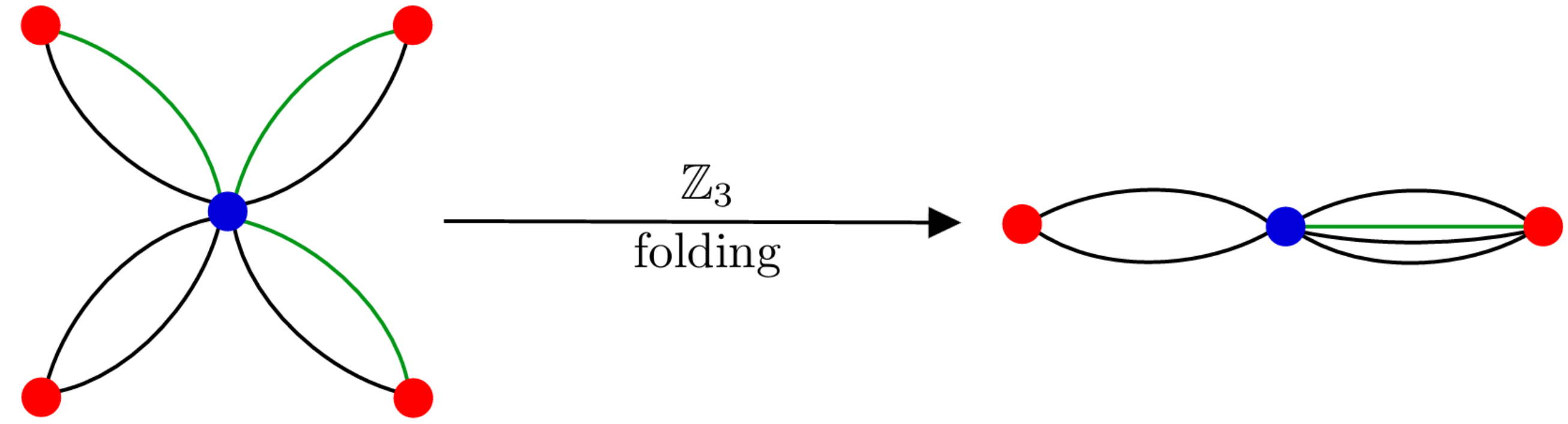}
\caption{ $  \mathbb{Z}_{3} $  folding of  the  tetra-quadric CY diagram. }
\label{F6}
\end{center}
\end{figure}

Here, we consider    the  $\mathbb{Z}_{3}$   folding  procedure.  This symmetry can  identify  three  red vertices associated with  three  different $ \mathbb{CP}^{1}$'s.    This  $\mathbb{Z}_{3}$  action  transforms these   vertices  and the associated  green links   as triplets    required by  the CY  dimension condition. This  procedure is  shown  in  Fig.(\ref{F6}).

This folding scenario leads to a digram which can  represent  a CY threefold in the  ambient projective space   $\mathbb{CP}^{1}\times\mathbb{CP}^{3}$.   The resulting  K\"{a}hler geometry involves two  dimensions less than  the natural one. This dimension reduction can be interpreted as a truncation  in the  black hole physics embedded in   the  M-theory on  the tetra-quadric CY manifold. It has  been observed that the  $ \mathbb{Z}_{3}$ actions are  guided  by  certain constraints that should  be imposed on  the  K\"{a}hler  and  the charge moduli spaces.
On such spaces, the  $\mathbb{Z}_{3}$  actions on the $t_i$ and $q_i$ quantities  can be accompanied by the following transformations
 \begin{eqnarray}
 &&\left(t_1, t_2, t_3, t_4\right) \to  2 \left(t_1, t_1, t_1,   \frac{1}{2}t_2\right)\\
&& \left(q_1, q_2, q_3, q_4\right)  \to     \frac{3}{2} \left(  q_1,  q_1,  q_1,\frac{2}{3}  q_2\right).
  \end{eqnarray}
Putting such  transformed   the  K\"{a}hler   moduli and  the charges in the   black hole scalar   potential Eq.(\ref{Vbh}), we recover  the scalar potential of 5D black holes obtained from M-theory on a   CICY  in   the ambient projective space   $\mathbb{CP}^{1}\times\mathbb{CP}^{3}$ as  reported in \cite{1}. In this way, 
the effective potential   of the M-theory on the   tetra-quadric CY reduces to 
\begin{equation}
V^{BH}_{eff}= \frac{G(t_1,t_2,q_1,q_2)}{{ T(t_1,t_2)}},
\end{equation}
where  we have used 
\begin{align*}
G(t_1,t_2,q_1,q_2)&= \frac{1}{12} q_2^2 \left(t_1^2+8 t_2 t_1+24 t_2^2\right)-\frac{1}{3} q_1 q_2 t_1^2+q_1^2 t_1^2\\
T(t_1,t_2) &= 1.
\end{align*} 
 Similar discussions can be elaborated  for black string potentials.   For these  solutions,  the  $ \mathbb{Z}_{3}  $ action  can be accompanied by the following transformations  on  $t_i$ and $p_i$    quantities 
 \begin{eqnarray}
 &&\left(t_1, t_2, t_3, t_4\right) \to \sqrt{6}  \left(2 t_1,2  t_1, 2t_1,  t_2\right)\\
&& \left(p_1, p_2, p_3, p_4\right)  \to    \frac{1}{8}\left(2p_1, 2p_1, 2p_1,  p_2\right).
  \end{eqnarray}
Handling  such transformed   $t_i$ and $p_i$,   we recover  the scalar  potential of 5D black  strings  obtained from M-theory on  the $ K3 $  fibration    in   the   projective space $\mathbb{CP}^{1}\times\mathbb{CP}^{3}$ investigated  in  \cite{1}.
 Using  similar techniques, the corresponding stringy
 effective potential  is found to be 
 \begin{eqnarray}
V^{BS}_{eff}= \frac{2}{3} t_1^2 \left(p_1^2 \left(t_1^2+8 t_2 t_1+24 t_2^2\right)+4 p_2 p_1 t_1^2+12 p_2^2 t_1^2\right).
\end{eqnarray}

 \subsection{$ \mathbb{Z}_{4}$  folding  procedure  in M-theory scenario}
\begin{figure}[!h]

\begin{center}
\includegraphics[scale=0.25]{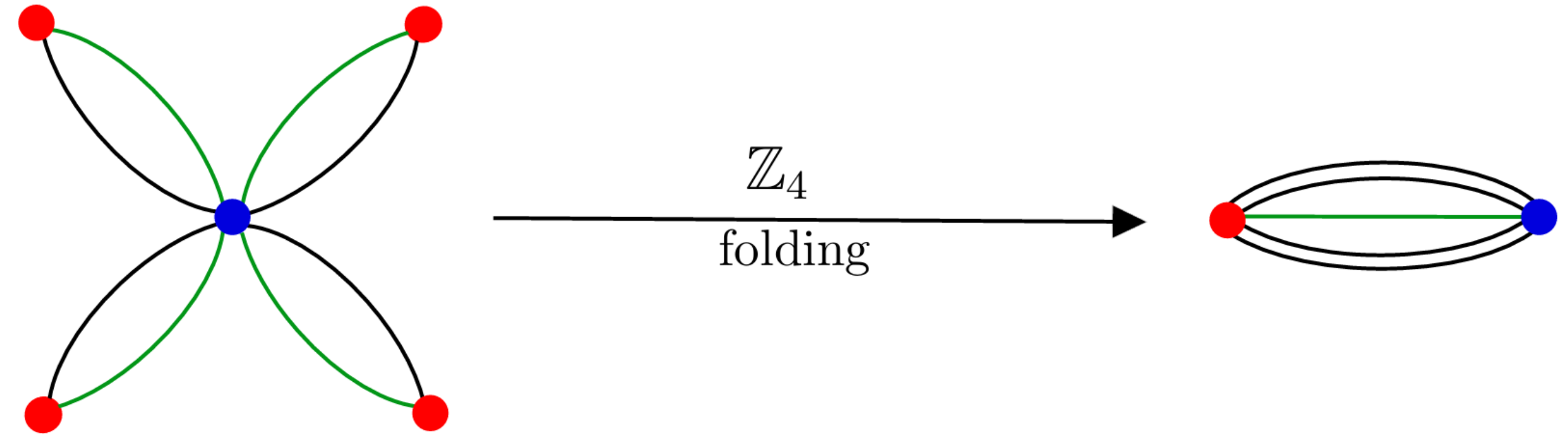}

\caption{ $  \mathbb{Z}_{4} $  folding of  the  tetra-quadric CY diagram. }
\label{F7}

\end{center}

\end{figure}

Finally,  we consider    the  $\mathbb{Z}_{4}$   folding  procedure.  This symmetry can  identify   four   red vertices corresponding to   four  $ \mathbb{CP}^{1}$'s.    
This  $\mathbb{Z}_{4}$  folding  action transforms  such  vertices and  the associated  green links   as quadruplets     required by  the CY  dimension condition. This  procedure is  shown  in  Fig.(\ref{F7}).

The present  folding scenario provides a   diagram  which can  represent  a  quintic in  the   projective space  $\mathbb{CP}^{4}$.   The resulting  K\"{a}hler geometry involves three   dimensions less than  the natural one. This dimension reduction can be interpreted as a truncation  in the  black hole physics embedded in  in M-theory on  the tetra-quadric CY manifold. It has  been observed that the  $ \mathbb{Z}_{4}$ action  can be  followed  by  certain constraints that should  be imposed on  the  K\"{a}hler  and charge moduli spaces of such a compactification.
On such spaces, the  $\mathbb{Z}_{4}$   folding action can be accompanied by the following transformations  on the $t_i$ and $q_i$ quantities  
 \begin{eqnarray}
 &&\left(t_1, t_2, t_3, t_4\right) \to 2  \left(t_1, t_1, t_1,   t_1\right)\\
&& \left(q_1, q_2, q_3, q_4\right)  \to    2 \left( q_1, q_1, q_1, q_1\right).
  \end{eqnarray}
Putting such  transformed   K\"{a}hler   moduli and charges in the   black hole scalar   potential Eq.(\ref{Vbh}), we  obtain  the scalar potential of 5D black holes obtained from M-theory on  the quintic CY manifold. In this way, 
the black hole  effective potential  is shown  to be
\begin{equation}
V^{BH}_{eff}= \frac{G(t_1,q_1)}{{ T(t_1)}},
\end{equation}
where  we have found 
\begin{align*}G(t_1,q_1)&= \frac{2}{3}q_1^2 t_1^2 \\
T(t_1) &= 1.
\end{align*} 
 Similar discussions can be  conducted   for 5D  black string potentials.   For such solutions,  the  $ \mathbb{Z}_{4}  $ action  can be accompanied by the following transformations  on  $t_i$ and $p_i$    quantities 
 \begin{eqnarray}
 &&\left(t_1, t_2, t_3, t_4\right) \to  \sqrt{6} \left( t_1, t_1, t_1, t_1\right)\\
&& \left(p_1, p_2, p_3, p_4\right)  \to  \frac{8}{5}  \left(p_1, p_1,p_1, p_1\right) .
  \end{eqnarray}
Taking such transformed   $t_i$ and $p_i$,   we  can obtain  the scalar  potential of 5D black  strings  obtained from M-theory on  the   quintic    in the projective space  $\mathbb{CP}^{4}$.
 Using  similar techniques, this  stringy 
 effective potential   is expressed as follows
\begin{eqnarray}
V^{BS}_{eff}=\frac{25}{6} p_1^2 t_1^4.
\end{eqnarray}

\section{Conclusion}
In this paper,  we have contributed to  the program of    the   construction  of  5D supersymmetric black branes from  the  M-theory  compactification.    Precisely,  we have   reconsidered  the study  of  5D black  holes and black  strings  using the  M-theory  compactification on  a special  CY manifold  called   tetra-quadric,  being   realized as  complete intersections  of   homogeneous   polynomials  in  the projective space $ \mathbb{CP}^{1}\times\mathbb{CP}^{1}\times\mathbb{CP}^{1}\times\mathbb{CP}^{1}$.   Using  colored graph theory and outer-automorphism  group action techniques,  we  have approached   the   tetra-quadric CY  diagram. We have shown that such  a graph is invariant under   the outer-automorphism  groups   $\mathbb{Z}_{2}$, $ \mathbb{Z}_{2} \times \mathbb{Z}_{2}$,   $ \mathbb{Z}_{3}$, and   $  \mathbb{Z}_{4}$ by identifying the permuted  red  vertices and green  legs.
 Using a  procedure referred to as folding,  we have recovered diagrams of certain CY manifolds. This feature has found a  place in the   construction  of  5D supersymmetric black holes and black  strings  using the  M-theory  compactification. Using a  procedure referred to as folding, we have shown  that  M-theory  black branes on the   tetra-quadric CY  manifold  can be  reduced to  known  compactifications  with lower  dimensional   K\"{a}hler moduli 
spaces.


\end{document}